# Comparison of the properties of GaN grown on complex Si-based structures

S.Q. Zhou, A. Vantomme [a)]

Instituut voor Kern- en Stralingsfysica, KULeuven, B-3001 Leuven, Belgium

B.S. Zhang, and H. Yang

State Key Laboratory on Integrated Optoelectronics, Institute of Semiconductors, Chinese Academy of Sciences, Beijing 100083, People's Republic of China

M.F. Wu

Department of Technical Physics, School of Physics, Peking University, Beijing 100871, People's Republic of China and Instituut voor Kern- en Stralingsfysica, KULeuven, B-3001 Leuven, Belgium

a) Corresponding author: Andre.Vantomme@fys.kuleuven.be

*Abstract*

With the aim of investigating the possible integration of opto-electronic devices, epitaxial GaN layers have been grown on Si(111) SOI and on Si/CoSi$_2$/Si(111) using metalorganic chemical vapor deposition. The samples are found to possess a highly oriented wurtzite structure, a uniform thickness and abrupt interfaces. The epitaxial orientation is determined as GaN(0001)//Si(111), GaN[11$\bar{2}$0]//Si[1$\bar{1}$0] and GaN[10$\bar{1}$0]//Si[$\bar{1}\bar{1}$2], and the GaN layer is tensily strained in the direction parallel to the interface. According to Rutherford backscattering/channeling spectrometry and (0002) rocking curves, the crystalline quality of GaN on Si(111) SOI is better than that of GaN on silicide. Room-temperature photoluminescence of GaN/SOI reveals a strong near-bandedge emission at 368 nm (3.37 eV) with an FWHM of 59 meV.

Due to the potential integration of microelectronics and optoelectronics, the growth of GaN on silicon substrates attracts a lot of attention [1]. A major challenge in the growth of GaN on Si(111) is the large mismatch of the in-plane thermal expansion coefficient ($5.59 \times 10^{-6}$ $K^{-1}$ for GaN and $2.6 \times 10^{-6}$ $K^{-1}$ for Si), which leads to cracking in the GaN layer when cooling the heterostructure from the growth temperature to room temperature. This problem has been largely solved by using a low temperature AlN buffer layer or interlayer [2], and based on this technique, crack-free blue light emitters have recently been demonstrated on Si(111) using InGaN/GaN multiple quantum wells structures [3].

Compared to the huge efforts of GaN growth on Si(111), by far less research has been performed on more complex technologically relevant Si-based structures, such as semiconductor-on-insulator (SOI) and silicides. Steckl *et al.* [4] reported GaN growth on a SiC(111) SOI structure, while Cao *et al.* [5] successfully grew GaN on Si(001) SOI. In this work, we report on the growth and properties of epitaxial GaN layers on Si(111) SOI and on Si/$CoSi_2$/Si(111) using metal-organic chemical vapor deposition (MOCVD).

The Si(111) SOI substrates were commercially prepared by SIMOX (separation by implanted oxygen) technology. The thickness of the silicon top-layer and $SiO_2$ buried layer is 200 nm and 360 nm, respectively. The $CoSi_2$ substrate was formed by ion beam synthesis using high dose ion implantation at elevated temperature followed by annealing [6]. In this work, 200 keV Co ions were implanted into Si(111) at 400°C to a fluence of $1.5 \times 10^{17}$ at./$cm^2$. The wafer was subsequently annealed at 1000°C in nitrogen for 20 min. The thickness of the silicon top-layer and the buried $CoSi_2$ layer is 90 nm and 60 nm, respectively. The GaN layers were grown directly onto these substrates without any buffer layer in an MOCVD reactor using $H_2$ as a carrier gas. Trimethylgallium (TMGa) and $NH_3$ were used as the precursors of Ga and N, respectively. Prior to GaN deposition, the substrates were heated to 1100ºC for 10 min under $H_2$ ambient to remove the native oxide from the surface. Subsequently the GaN layers were deposited at 1100ºC.

Before and after GaN growth, the substrates were analyzed by Rutherford backscattering/channeling spectrometry (RBS/C) using a collimated 1.57 MeV $He^+$ beam. $\chi_{min}$, the channeling minimum yield, which is the ratio of the backscattering yield when the impinging beam is aligned to a crystallographic axis ($Y_A$) to that for a random beam incidence ($Y_R$), is a measure of the crystalline quality of the film. Fig. 1 shows the RBS/C spectra of both samples before and after GaN deposition, along with simulations. For $CoSi_2$ [Fig. 1 (a)] and Si(111) SOI [Fig. 1 (b)], the $\chi_{min}$ for the Si top-layer is 11% and 2.1%, respectively, revealing that the crystalline quality of the Si top-layer, which serves as a template for further GaN growth, is better for the Si(111) SOI structure than for the $CoSi_2$ substrates. The interfaces in both heterostructures are very abrupt within the sensitivity of backscattering spectrometry. The interfaces between the nitrides and Si substrates after GaN growth [Fig.1 (c) and (d)] are abrupt as well, and the $\chi_{min}$ of GaN is 2.3% and 5.2% for the SOI and $CoSi_2$ substrates, respectively. Hence, whereas the crystalline quality of GaN/SOI approaches that of the best layers grown on sapphire where the typical $\chi_{min}$ value is 1~2%, layers grown onto buried silicides contain a larger degree of lattice imperfections. The $\chi_{min}$ values are listed in table I for comparison. As observed, the quality of the Si top-layer directly influences that of the GaN epilayer. Scanning cross section images of the samples (not shown) clearly confirm that the specimens have a well-defined multilayer structure with uniform thickness and abrupt interfaces.

High resolution X-ray diffraction was used to investigate the crystalline structure of the GaN layers. The FWHM of the X-ray rocking curve for the GaN(0002) diffraction is 22.8 arcmin for GaN/Si(111) SOI and 46.8 arcmin for GaN/CoSi$_2$, which confirms the better quality of GaN/SOI.

The epitaxial relationship of the nitrides was revealed by X-ray diffraction reciprocal space mappings of the GaN films and substrates (Fig. 2) and by in-plane Φ-scans of the GaN(11$\bar{2}$2) and Si($\bar{1}$13) planes (not shown). The Si($\bar{1}$13) diffraction reveals a sixfold symmetry at the same azimuthal positions as the GaN(11$\bar{2}$2) planes. Fig. 2 shows the reciprocal spacing mapping of GaN on SOI from the (111) reciprocal lattice point to the (004) point. The X-ray was incident along the Si[1$\bar{1}$0] direction. By inclining the sample from -1° to +59°, the substrate related diffractions of (113) and (004) planes in the Si[1$\bar{1}$0] zone are observed at an inclination angle of 29.5° and 54.8°, respectively. At the same time, the GaN related diffractions from (10$\bar{1}$2) and (10$\bar{1}$3) in the GaN[11$\bar{2}$0] zone were observed at an inclination angle of 43.2° and 32.0°, respectively. From this pattern, the epitaxial orientation is determined to be GaN(0001)//Si(111), GaN[11$\bar{2}$0]//Si[1$\bar{1}$0] and GaN [10$\bar{1}$0]//Si[$\bar{1}\bar{1}$2].

Based on the abovedetermined epitaxial relationship, the bulk lattice parameters of GaN and the Si substrate are $a_{GaN}$ = 0.3189 nm, and $a_{Si(111)}$ = 0.3840 nm, resulting in a large lattice mismatch of −16.9%. The lattice constants of the epitaxial GaN layers were precisely determined by XRD symmetric and skew symmetric θ-2θ scans. Fig. 3 (a) and (b) shows the GaN(0002), the GaN(0004) and the substrate Si(111) and Si(222) reflection peaks. Within the instrumental detection limit, no other nitride peaks were observed, indicating that the GaN epitaxial layers solely consist of the wurtzite phase, and the nitrides are perfectly c-axis oriented. Taking the Si(111) peak as the reference, the Bragg angle of GaN(0002) can be accurately determined. In order to eliminate the possible inter-layer tilt, a second θ-2θ scan was carried out after rotating the sample by 180˚. According to the Bragg equation $2d\sin\theta = \lambda$ where $d$ is the lattice spacing, $\theta$ is the Bragg angle and $\lambda$ is the incident X-ray wavelength, the c lattice constants for GaN/Si(111) SOI and GaN/CoSi$_2$ are determined to be 0.5179 nm and 0.5182 nm, respectively. To determine the in-plane lattice constant, a θ-2θ scan from GaN(11$\bar{2}$2) and Si($\bar{1}$13) was performed in a skew symmetric geometry by tilting the sample by 58.5˚ [Fig. 3 (c) and (d)]. Using the abovementioned method, the in-plane lattice constants are determined to be 0.3190 nm and 0.3196 nm for GaN layer on Si(111) SOI and on CoSi$_2$, respectively. Using the obtained lattice constants, the residual elastic strain can be evaluated by the following equations [7]

$$e^\perp = \frac{c_{epi} - c_b}{c_b} \text{, and } e^\parallel = \frac{a_{epi} - a_b}{a_b}$$

where $e^\perp$ and $e^\parallel$ are the elastic strain in perpendicular and parallel directions, respectively, $c_{epi}$, $a_{epi}$, $c_b$, and $a_b$ are the lattice constants for epilayer and bulk material. The GaN layer is found to be tensily and compressively strained in the in-plane and out-of-plane direction, respectively, which is in agreement with the negative lattice mismatch between GaN and Si(111). The calculated elastic strain values are listed in table 2. Within the experimental error, the in-plane strain for GaN/SOI is approximately zero, which is attributed to the stress absorption in the compliant Si layer on top of the *amorphous* $SiO_2$ [8]. In contrast, for silicides, the top Si layer is not compliant since the $Si/CoSi_2$ heterostructure is *epitaxially* grown onto the Si(111) wafer. Although RBS/C is a depth sensitive technique to determine the elastic strain [9] (in contrast to XRD, where the information is typically averaged over a depth of several μm), it is impossible to investigate the strain alteration in the thin top Si layer. Indeed, due to the different lattice structure of GaN and Si, the channelling effect in off-normal axes of the buried Si layer virtually disappears, making strain determination impossible.

The optical properties of the GaN layer grown on Si(111) SOI were investigated by photoluminescence (PL) at room temperature, using a He-Cd laser with a wavelength of 325 nm (Fig. 4). A dominant peak with an FWHM of 59 meV is observed at 368 nm (3.37 eV), corresponding to the near-bandedge emission in a wurtzite GaN epitaxial layer. The FWHM is smaller than the value of 140-150 meV reported for GaN grown on SiC(111) SOI [4], and comparable with the value of 33 meV for high quality GaN layers grown directly on Si(111) [10]. A rather broad yellow band appears near 550 nm (2.25 eV). The intensity ratio of the near-bandedge to the yellow band is approximately 5.6, which is larger than the value given in reference [4] (estimated to be around 1.4).

In conclusion, epitaxial GaN films have been successfully formed on complex Si-based structures of Si(111) SOI and silicides. The multi-layered samples exhibit a good crystalline quality, uniform thickness and abrupt interfaces. At room temperature, the near-bandedge emission is observed in the GaN layer on Si(111) SOI by photoluminescence while the yellow band is rather weak. The epitaxial orientation is determined to be GaN(0001)//Si(111), GaN[11$\bar{2}$0]//Si[1$\bar{1}$0] and GaN [10$\bar{1}$0]//Si[$\bar{1}\bar{1}$2], and the nitride layer is tensily and compressively strained in the in-plane and out-of-plane direction, respectively. It is found that the crystalline quality of GaN on Si(111) SOI is better than that of GaN on silicide. The crystalline quality of the GaN layer appears to be directly influenced by the Si top-layer.

The authors are grateful to Dr. X. P. Ye for the scanning electron microscopy measurement. This work was supported by the Fund for Scientific Research, Flanders (FWO), the Concerted Action of the KULeuven (GOA/2004/02), the Inter-University Attraction Pole (IUAP P5/1), the Bilateral Cooperation between China and Flanders (BIL 02-02), and by the National Natural Science Foundation of China under the grant No. 10075072.

Reference:

[1] A. Krost, and A. Dadgar, Mat. Sci. & Eng. **B93**, 77 (2002) and references therein.

[2] A. Dadgar, J. Bläsing, A. Diez, A. Alam, M. Heuken and A. Krost, Jpn. J. Appl. Phys. **39**, L1183 (2000).

[3] A. Dadgar, M. Poschenrieder, J. Bläsing, K. Fehse, A. Diez, and A. Krost, Appl. Phys. Lett. **80**, 3670 (2002).

[4] A.J. Steckl, J. Devrajan, C. Tran, and R.A. Stall, Appl. Phys. Lett. **69,** 2264 (1996).

[5] J. Cao, D. Pavlidis, A. Eisenbach, A. Philippe, C. Bru-Chevallier, and G. Guillot, Appl. Phys. Lett. **71,** 3880 (1997).

[6] A. Vantomme, M. F. Wu, I. Dézsi, G. Langouche, K. Maex and J. Vanhellemont, Mat. Sci. & Eng. **B4**, 157 (1989).

[7] M. F. Wu, A. Vantomme, S. M. Hogg, G. Langouche, W. Van der Stricht, K. Jacobs, and I. Moerman, Appl. Phys. Lett. **74**, 365 (1999).

[8] J. Cao, D. Pavlidis, Y. Park, J. Singh, and A. Eisenbach, J. Appl. Phys. **83,** 3829 (1998).

[9] M. F. Wu, C. Chen, D. Zhu, S.Q. Zhou, A. Vantomme, G. Langouche, B.S. Zhang and H. Yang, Appl. Phys. Lett. **80**, 4130 (2002).

[10] M. H. Kim, Y. C. Bang, N. M. Park, C. J. Choi, T. Y. Seong, and S. J Park, Appl. Phys. Lett. **78**, 2858 (2001).

Table I Crystalline quality of Si top-layer and GaN epilayer (see text for explanation).

|  | $\chi_{min}$ Si top layer | $\chi_{min}$ GaN | FWHM GaN(0002) |
|---|---|---|---|
| GaN/SOI | 2.1% | 2.3% | 22.8 arcmin |
| GaN/CoSi$_2$ | 11% | 5.2% | 46.8 arcmin |

Table II Lattice constants and elastic strain in the GaN layers

|  | $c_{epi}$ (nm) | $a_{epi}$ (nm) | $e^{\perp}$ (%) | $e^{\parallel}$ (%) |
|---|---|---|---|---|
| GaN/SOI | 0.5179±0.0003 | 0.3190±0.0003 | - 0.12±0.06 | + 0.03±0.09 |
| GaN/CoSi$_2$ | 0.5181±0.0003 | 0.3196±0.0003 | - 0.08±0.06 | + 0.22±0.09 |

Fig captions

Fig. 1, Random (o), aligned (+) and simulated (−) backscattering spectra using a scattering angle of 170° (a) buried $CoSi_2$ in Si(111), (b) Si(111) SOI, (c) GaN on $CoSi_2$, and (d) GaN on Si(111) SOI.

Fig. 2, Reciprocal space mapping of the GaN/SOI sample in the Si[$\bar{1}$10] zone. The diffractions from Si(113), Si(004), GaN(10$\bar{1}$2), and GaN(10$\bar{1}$3) are indicated. The resulting epitaxial relationship is GaN(0001)//Si(111), GaN[11$\bar{2}$0]//Si[1$\bar{1}$0] and GaN[10$\bar{1}$0]//Si[$\bar{1}\bar{1}$2].

Fig. 3, 2θ-θ X-ray diffraction patterns: GaN (0002) (0004) and Si(111) (a) GaN on $CoSi_2$ and (b) GaN on Si(111) SOI, and GaN(11$\bar{2}$2) and Si($\bar{1}$13) in skew symmetric geometry (c) GaN on $CoSi_2$ and (d) GaN on Si(111) SOI. Neither other phases nor any other growth direction were detected.

Fig. 4, Room temperature photoluminescence spectrum of GaN on Si(111) SOI.

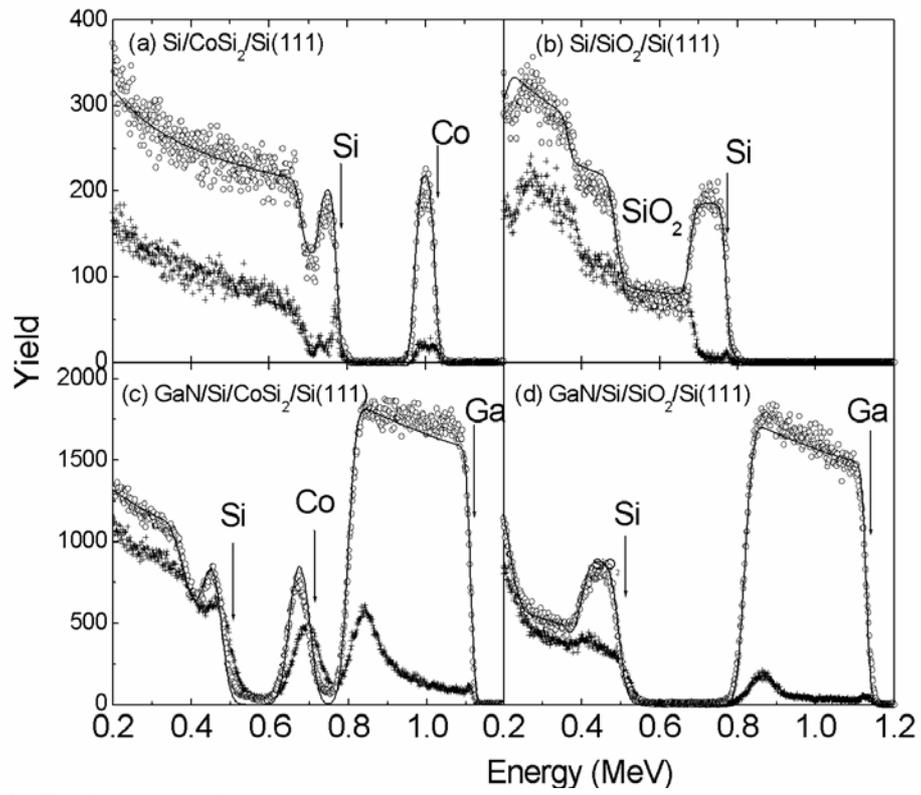

S. Q. Zhou, Fig.1

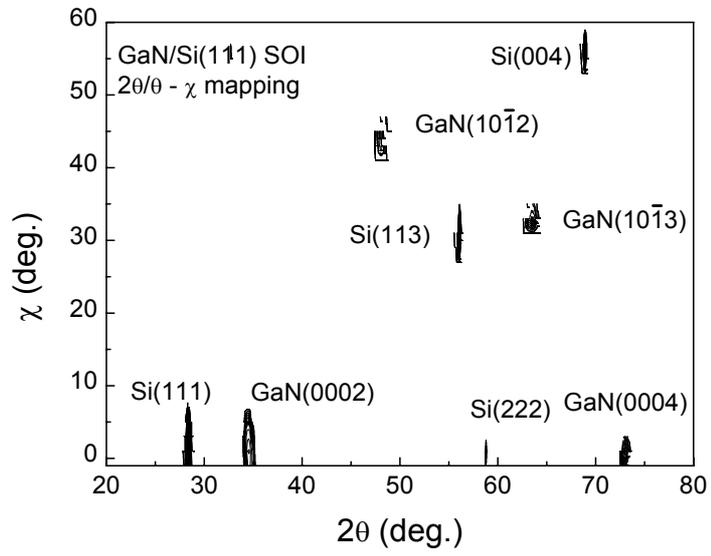

S. Q. Zhou, Fig. 2

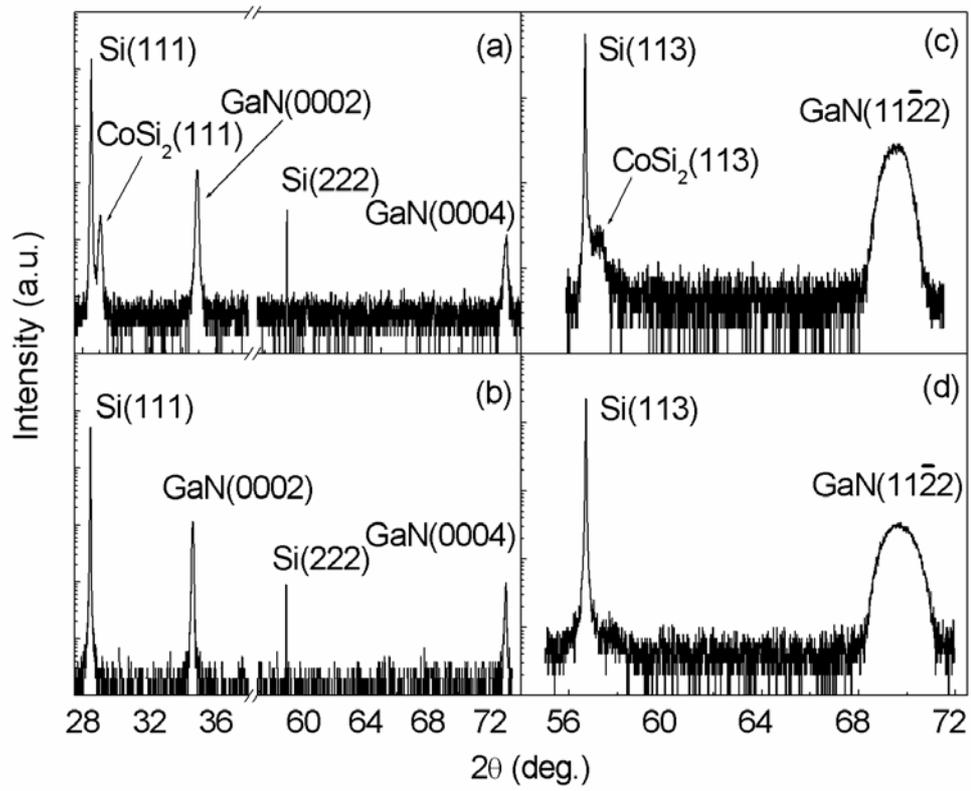

S. Q. Zhou, Fig. 3

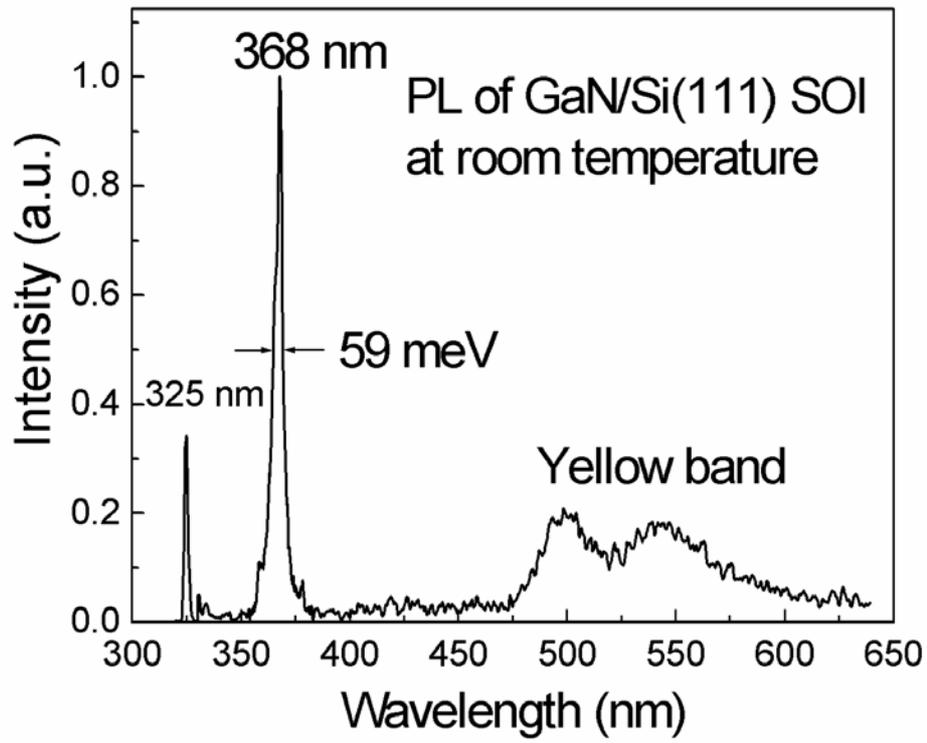

S. Q. Zhou, Fig.4